\documentstyle[epsfig,longtable]{aipproc}

\begin{document}

\def\gsim{\mathrel{\raise.3ex\hbox{$>$}\mkern-14mu
	     \lower0.6ex\hbox{$\sim$}}}
\def\lsim{\mathrel{\raise.3ex\hbox{$<$}\mkern-14mu
	     \lower0.6ex\hbox{$\sim$}}}

\title{Modeling the Time Variability of Black Hole Candidates}
 
\author{Demosthenes Kazanas$^*$, Xin-Min Hua$^{*\dagger}$}
\address{$^*$NASA/GSFC, Code 661, Greenbelt, MD 20771\\
$^{\dagger}$Universities Space Research Association }

\maketitle

\begin{abstract}
We present model light curves, Power Spectral Densities (PSD) and 
time lags of accreting Black Hole Candidates (BHC)  based on a 
recent model of these sources.  According to our model the observed
variability is due, to a large extent, to the stochastic nature of 
Comptonization, the basic process responsible also for the emission of 
high energy radiation. Our additional assumption is 
that  the Comptonization process takes place in an extended but 
non-uniform atmosphere around the compact object. Our model reproduces 
the observed light curves well, in that it provides a good 
fit to the PSDs and the overall light curve morphology, 
indicating, in accordance with observation, that most of the power 
resides at time scales $\gsim$ a few seconds while at the same time 
one can distinguish shots of a few msec in duration. We suggest that 
refinement of this type of model along with spectral and phase lag 
information can be used to probe the structure of this class of 
high energy sources.
\end{abstract}

\section*{Introduction}

The study of the physics of accretion onto compact objects (neutron 
stars and black holes) involves length scales much too small to 
be resolved by the current or the forseable future technology.  As such, 
this study is conducted mainly by the theoretical 
interpretation of spectral and temporal observations of these systems. 
In this respect, spectral analysis presents the first line of attack 
in unfolding their physical properties. For the specific systems and 
in particular Black Hole Candidate (BHC) sources, a multitude of 
observations have indicated that their energy spectra can be fitted 
very well by Comptonization of soft photons by hot electrons. 
It is thus generally agreed that Comptonization 
is the process by which the high energy ($\gsim 2-100$  keV) spectra 
of these sources are formed. 
Thus, while the issue of the detailed dynamics of accretion onto the 
compact object is still not resolved, the assumption of the presence 
of a thermal distribution of hot electrons has proven sufficient to 
produce models which successfully fit the spectra of the emerging high 
energy radiation. Spectral fitting, has been hence used as 
a probe of the dynamics of accretion onto the compact object.  

However, the Comptonization spectra, which depend roughly on the 
product of the electron temperature and the photon escape probability 
from the scattering cloud, do not provide in and of themselves a 
measure of the scale of the system. As such they cannot provide any 
clues about the dynamics of accretion of the hot gas onto the compact 
object, let alone its parameters (density, velocity) as a 
function of radius, necessary to determine these dynamics. One needs, 
in addition, time variability information. 

While simple minded dynamics anticipate most of the variability
power to be associated with the dynamical time scales of  the 
last few gravitational radii (msec for galactic sources), the observed 
PSD indicate that the variability power resides at time scales $\gsim 1$ 
sec, far removed from the ``natural" ones. At the same time, however, the
observed light curves do exhibit the presence of ``flares" with the 
expected (a few msec) rise times (see e.g. \cite{meek84}), even though the 
latter make an insignificant contribution to the PSD. 
It is therefore thought that the observed PSDs are due to a 
modulation of the accretion rate onto the compact object, the result 
of a kind of self-organized-criticality for the dynamics of 
accretion \cite{TMN95}.
 
However, the Comptonization process, provides a much more 
refined probe of variability than the PSD: Because, on the
average, the energy of the escaping photons increases 
with their residence time in the scattering medium, the 
hard photon light curves lag with respect to those of softer photons 
by amounts which depend on the photon scattering time. Therefore,
if the high energy radiation is emitted from a compact region near the 
black hole, the resulting lags $\Delta t$ should be independent of the 
Fourier components of the time variation and $\Delta t \simeq$ msec. 

These lags were sought for and Fourier analyzed in the light curves 
of Cyg X-1 \cite{{miy88},{miy91},{cui97b}}, the X-ray 
transient source J0422+32 \cite{gro97} and the galactic source GRS 
1758-258 \cite{smith97}. Surprizingly, rather than being constant 
($\simeq$ msec), as expected, their magnitude was found to increase 
linearly with the Fourier period $P$, from $\lsim $0.001 sec at 
$P \simeq 0.05$ sec to roughly $0.1-1$ sec for $P \simeq 10$ sec. 
Their magnitude and in particular their 
$P$-dependence, exclude the possibility that the observed PSD are due 
to modulation of the accretion rate and strongly indicate scattering in 
a very extended non-uniform medium, whose density profile reflects on 
the functional dependence of the lags on $P$ as suggested by 
\cite{{KHT},{HKT},{HKC}}. 

As suggested originally in \cite{KHT} and discussed in more detail in 
\cite{{HKT},{HKC},{HKCb}}, the issue of time lags can be 
resolved if the Comptonization takes place in a very extended ($\lsim 
10^{11}$ cm) non-uniform hot plasma with density profile $n(r) \propto 
r^{-p}, ~p \simeq 1$, the range in $P$ reflecting the range in radius 
over which this configuration extends. It was further discussed there that 
the dependence of lags on $P$ relates to the value of the index $p$ and
it can be used to probe the density structure of the high 
energy emitting region. Herein we discuss and present models on the 
X-ray light curves associated with this specific model of high energy
emission from accreting compact objects.

\section*{Modeling the Light Curves}

The fundamental tenet of our model is that the observed variability is
due largely to the stochastic nature of Compton scattering. As discussed
in \cite{KHT}, the response of the extended atmosphere to a 
$\delta$-function injection of soft photons in its center, has a power
law (rather than exponential as is the case for a uniform medium) 
time dependence. Our model light curves consist of an incoherent 
sum of such responses, the result of random injection of soft photons at
given rate determined by the dynamics of the smallest radii, i.e.
$$F(t) = \sum_{i=1}^N Q_i g(t - \tau_i)  \eqno(1)$$
\noindent where $g(t)$ is the response function of the atmosphere, 
$\tau_i$ are the arrival times of the individual
shots and $Q_i$ their amplitudes. To avoid introducing information
in the distribution of $Q_i$'s and $\tau_i$'s we assume that the 
$Q_i$'s are constant while the $\tau_i$'s are randomly distributed 
about a mean value. 

To perform the sum of Eq.(1) we produce an analytic fit to the 
function $g(t)$ whose form has been computed using the Monte-Carlo 
code developed by one of us (XMH). The form used in fitting them is 
the following:
$$g(t)= \cases{A_1 (1-B_1 x^b)x^{\gamma}, &if $x \equiv t/t_0 \le 1$;\cr
                A_2 (1+B_2x^{-b})x^{\alpha -1}e^{-(xt_0/\beta)^3}, 
&if $x \equiv t/t_0 > 1$, \cr } \eqno(2)$$
\begin{figure} 
\centerline{\epsfig{file=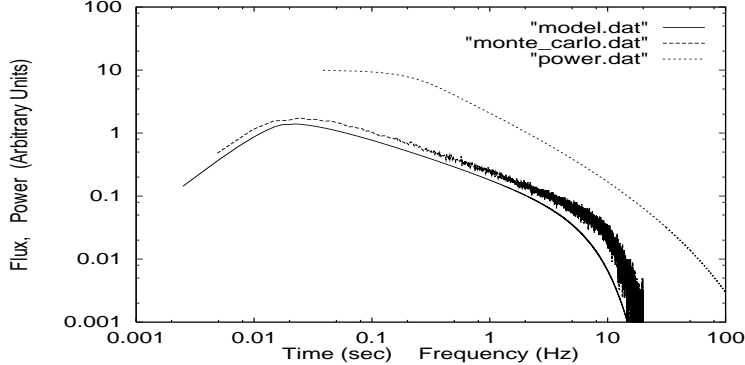,width=4.0in,height=1.9in}}
\vspace{10pt}
\caption{
The shot profiles as computed by the Monte-Carlo (dashed line) and their
analytical fits (solid line), along with the corresponding PSD (dotted line).
}\label{Figure 1}
\end{figure}

The model and computed response functions are given in Fig. 1. The 
values of the parameters corresponding to the specific figure are 
$t_0 = 0.02,~\beta = 10$ sec, and $\alpha \simeq 0.50, ~b=1,~\gamma=
1.5$. The agreement between them is excellent. 

The PSD of the light curve of Fig. 2, computed using Eq. (1), due 
to the independence of the shots, is simply the PSD of an 
individual shot, also shown in Fig. 1. This PSD, 
besides the segment $\propto \omega^{-2 \alpha}$, reflecting 
the $t^{\alpha -1}$ part of $g(t)$, exhibits also a flattening at 
$\omega \lsim 1/\beta$ and a cut-off at $\omega \gsim
10$ Hz, due respectively to the cut-off and rising parts of the shot.

\begin{figure} 
\centerline{\epsfig{file=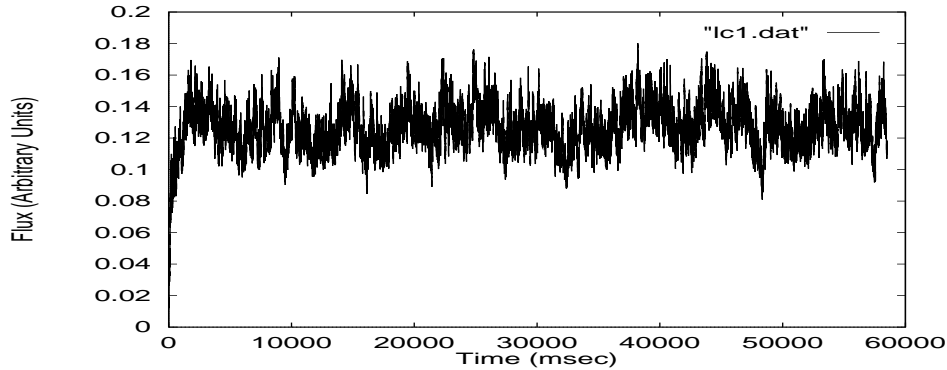,width=5.0in,height=1.9in}}
\vspace{10pt}
\caption{
A light curve computed using the prescription given in 
the text with $t_0 = 0.001$, $\beta = 1$ sec and $\alpha = 0.5$.
}\label{Figure 2}
\end{figure}

In figure 2 we present a light curve given by the above prescription.
The corrsponding values of the shot parameters are $\alpha = 0.5$, and
$t_0 = 0.001,~ \beta = 1$ sec. The initial rising segment lasts 
$\simeq \beta$ sec, i.e. the cut-off  time of the $g(t)$ function 
and it is only a turn-on transient. Eye inspection reveals the presence
of shots with a broad range of rising times and durations, down to $t_0$, 
and as long as a few seconds, even though only a single type of shot has been 
used in the model. These shots of varying duration are simply the
result of the random sum of shots with power-law tails. One should
note that the specific form of $g(t)$ cannot be discerned in 
their random superposition and one might naively consider the light 
curve as a sum of exponential shots with a distribution of timescales.
However, the power law structure of the underlying shots is apparent
in the PSD of the light curves.  At the same time, the fact that the 
time lags, as computed within this simple model \cite{{HKC},{HKCb}},  
fit the observations in such straigtforward fashion, argues strongly 
in support of the above interpretation.


\begin{references}




\bibitem{meek84}
Meekins, J. F. et al. 1984, ApJ, 278, 288

\bibitem{TMN95}
Takeuchi, M., Mineshige, S \& Negoro, H. 1995, PASJ, 47, 617

\bibitem{miy88}
Miyamoto, S. et al., 1988, Nature, 336, 450

 
\bibitem{miy91}
Miyamoto, S. et al., 1991, ApJ, 383, 784

\bibitem{cui97b}
Cui, W. et al. 1997,  ApJ, 484, 383

\bibitem{gro97}
Grove, J.E. et al. 1997, Proc. 4th Compton Symposium, 
AIP Conf. Proc. 410, p. 122

\bibitem{smith97}
Smith, D. M. et al. 1997, ApJL, 489, L51

\bibitem{KHT}
Kazanas, D., Hua, X.-M. \& Titarchuk, L. 1997 ApJ, 480, 735

\bibitem{HKT}
Hua, X.-M., Kazanas, D. \& Titarchuk, L. 1997 ApJ, 482, L57


\bibitem{HKC}
Hua, X.-M., Kazanas, D. \& Cui, W. 1997, ApJ Submitted (astro-ph/9710184)

\bibitem{HKCb}
Hua, X.-M., Kazanas, D. \& Cui, W. 1997, these Proceedings


\end{references}
\end{document}